\documentstyle[nato]{crckapb} 

\input{psfig}

\newcommand{\bp}{$\beta\:$Pictoris}

\newcommand{\ci}{\mbox{C~{\sc i}}}

\def\ga{\mathrel{\mathchoice {\vcenter{\offinterlineskip\halign{\hfil
$\displaystyle##$\hfil\cr>\cr\sim\cr}}}
{\vcenter{\offinterlineskip\halign{\hfil$\textstyle##$\hfil\cr
>\cr\sim\cr}}}
{\vcenter{\offinterlineskip\halign{\hfil$\scriptstyle##$\hfil\cr
>\cr\sim\cr}}}
{\vcenter{\offinterlineskip\halign{\hfil$\scriptscriptstyle##$\hfil\cr
>\cr\sim\cr}}}}}

\begin{opening}
\title{CIRCUMSTELLAR DISKS AND OUTER PLANET FORMATION}
\author{A. LECAVELIER DES ETANGS}
\institute{
Institut d'Astrophysique de Paris\\
98 Bld Arago, F-75014 Paris, France}
\end{opening}

\begin{document}

\begin{abstract}

The dust disk around \bp\ must be produced by collision or by evaporation of
orbiting Kuiper belt-like objects. Here we present the 
Orbi\-ting-Evaporating-Bodies (OEB) scenario in which the disk is 
a gigantic multi-cometary tail supplied by slowly evaporating bodies
like Chiron.
If dust is produced by evaporation, a planet with an eccentric orbit 
can explain the 
observed asymmetry of the disk, because the periastron distribution of the 
parent bodies are then expected to be non-axisymmetric.
We investigate the consequence for the Kuiper belt-like objects of 
the formation and the migration of an outer planet
like Neptune in Fern\'andez's scheme (1982). 
We find that bodies trapped in
resonance with a migrating planet 
can significantly evaporate, producing a \bp -like disk
with similar characteristics like opening angle and asymmetry.

We thus show that the \bp\ disk can be a transient phenomenon. 
The circumstellar
disks around main sequence stars can be the signature of the present
formation and migration of outer planets.

\end{abstract}

\section{Introduction}
\label{intro}

The infrared excess
Vega-like stars and their circumstellar dusty 
environment have been discovered 
by IRAS in the 80's (Aumann et al. 1984).
Among these infrared excess stars, \bp\ has a very peculiar status
because images have shown that the dust shell is in fact
a disk seen edge-on from the Earth (Smith \& Terrile 1984) and 
have given unique information on the dust distribution.
The disk morphology and the inferred spatial distribution of the dust
have been carried out in great details (Artymowicz et al. 1989, 
Kalas \& Jewitt 1995). The morphological properties 
can be summarized as follows
(see Lecavelier des Etangs et al., 1996, hereafter LVF):
First, the gradient of the scattered light follows a relatively well-known 
power law.
But the slope of this power law changes at about 120~AU from the star 
(Goli\-mowski et al. 1993).
Second, the disk has an inner hole with a central part relatively clear of dust
(Lagage \& Pantin 1994).
In the third dimension, the disk is a ``wedge'' disk: the thickness increases 
with radius (Backman \& Paresce 1993).
More surprisingly, the disk is not symmetric with
one branch brighter than the other
(see details in Kalas \& Jewitt, 1995).
Finally, the inner part of the disk ($\sim 40$~AU) seems to be warped. This
warp has been well-explained by Mouillet et al. (1997) as due to an inclined
planet inside the disk.

As the dust particle life-time is shorter than the age of the system,
one must consider that the observed dust is continuously resupplied
(Backman \& Paresce 1993). In order to explain the origin
of the dust in the \bp\ disk and these well-known
morphological properties, we have proposed 
the {\em Orbiting-Evaporating-Bodies} model (hereafter OEB)
(see LVF).

After a brief summary of the OEB scenario (Sect.~\ref{OEB et al}), 
we present its consequences on the explanation for
the presence of CO (Sect.~\ref{The Carbon Bearing Gas}) and
the asymmetry (Sect.~\ref{The Asymmetry Problem}).
Then, we will see that the \bp\ disk can be a natural
consequence of the formation of Neptune-like outer planets
(Sect.~\ref{migrating planet}).

\section{Summary of the Orbiting-Evaporating-Bodies Scenario}
\label{OEB et al}

The observed dust is continuously resupplied.
Two mechanisms can produce dust in this low density disk: collision
or evaporation of kilometer-sized parent bodies. 
In both cases, because of the radiation pressure, the 
particles ejected from the parent bodies follow very eccentric orbits
whose eccentricity is related to the grain size (Burns et al. 1979). 
If we assume a zero-order model of a narrow ring of bodies producing dust,
the particles are then distributed on a disk-like structure presenting
three morphological similarities with the \bp\ disk.
First, the central region of the disk is empty of dust,
its limit corresponds to the inner radius of the parent bodies' orbits.
Second, this zero-order model disk is open because the distribution
of the particles inclinations are the same as that of the parent bodies. 
Last, the dust density is decreasing with the distance to the star,
moreover this density distribution follows a power law.
Consequently, if seen edge-on from the Earth, 
the radial brightness profile along the mid-plane of this disk
follows also a power law: $F(r)\propto r^{-\alpha}$, with $\alpha\sim 5$ 
(LVF).

We can conclude that a ring of parent bodies on circular orbits
can naturally produce a disk with an inner hole, 
which is open, and if seen edge-on,
the scattered light distribution follows a power law.

But the slope of this zero order model is steeper than the observed 
slope of the power law in the \bp\ disk 
($\alpha \sim 4$, Kalas \& Jewitt 1995).
To explain this distribution,
it is possible that the dust is produced by collisions of Kuiper belt-like
objects spread in a wide range of distances.
But, an alternate solution is also possible in keeping  
the assumption that
the parent bodies remain in a narrow ring close to the star.
In that case, a large quantity of small particles is needed,
because these particles have larger apoastron and can explain the
less steep slope in the power law.
These small particles can typically be produced by the evaporation
of parent bodies of size $\ga 10$km and located at large distances.
Indeed, if the evaporation rate is small enough,
there is a cut-off on the maximal size of the particles which 
can be ejected from 
the bodies gravitational field by the evaporating gas.
This slow evaporation and peculiar particle
size distribution is observed in the Solar System around Chiron
(Elliot et al. 1995, Meech et al. 1997). 

Several arguments are in favor of this scenario
in the case of the \bp\ disk.
First, it is obviously easy to explain any asymmetry even at 
large distances, because a planet in the inner disk (on eccentric orbit) 
can have influence on the distribution of nearby parent bodies, and
this non-axisymmetric distribution is projected outward
by the particle on very eccentric orbits.
Of course the CO/dust ratio is one of the arguments which is in
favour of the OEB scenario (see Sect.~\ref{The Carbon Bearing Gas}).
Finally, the connection between the inner radius of the disk and
evaporation limit is a direct consequence of this scenario because
the periastron distances of the particles are similar to the periastron
of the parent bodies. Any hypothetical planet at this limit is no
more needed to explain the presence of the inner void in the disk.

\section{The Carbon Bearing Gas: a Clue to Evaporation}
\label{The Carbon Bearing Gas}

\subsection{A Needed Source of CO}
\label{CO and ci}

An important characteristic of the \bp\ gaseous disk is
the presence of cold CO and \ci\ (Vidal-Madjar et
al. 1994).
CO is cold with a typical temperature of less than 30~K which corresponds
to the temperature of CO-evaporation; for instance, with an albedo of 0.5 
this temperature corresponds to an evaporating body located between 100 and 
200~AU.
CO and \ci\ are destroyed by UV interstellar photons 
and have lifetime shorter than the star age 
($t_{CO}\sim t_{CI}\sim$200~years). 
A permanent replenishment mechanism must exist.  
To estimate the supplying rate of CO, one must assume a cloud 
geometry which gives
the connection between the observed column density and the total
CO mass. Assuming a disk geometry with an opening angle similar to the 
dust disk 
($\theta=7$~degrees), and a characteristic distance 
given by the CO temperature ($r_0=150$~AU), we get a mass of CO:
$%\begin{equation}
M_{CO}\approx 4\pi \theta \mu_{CO} N_{CO} r_0^2 \approx 7\times 10^{20} {\rm kg}
$, where $\mu_{CO}$ is the molecular weight and 
$N_{CO}=2\times 10^{15}$~cm$^{-2}$ is the column density of CO.
Then, the known photodissociation rate of CO, $\tau_{CO}=2\cdot 10^{-10}$s$^{-1}$ 
(Van Dishoeck \& Black, 1988)
gives a relation between the total CO mass 
and the corresponding supplying rate. We obtain
$%\begin{equation}
\dot{M_{CO}} = M_{CO}\tau_{CO} \approx 10^{11} {\rm kg\ s}^{-1}
$.%\end{equation}

We can also estimate the needed supplying rate of dust
$\dot{M_d} = M_d / t_d \approx 10^{11} {\rm kg\ s}^{-1}$,
where $M_d$ is the mass of the dust disk ($M_d\sim 10^{23}$kg), 
$t_d$ is the dust life-time ($t_d\approx10^4$yr).
It is very interesting to see that the dust/CO supplying rate is 
consequently $\dot{M_d}/\dot{M_{CO}}\approx 1$.
This very similar to
the ratio in the material supplied by evaporation in the solar 
system.
This provide an {\em independent}
evidence that the \bp\ dust disk can be supplied by 
{\em Orbiting-Evaporating-Bodies}.

We can now estimate the number of bodies producing this CO.
If we take an evaporation rate of CO 
per body $Z_{\rm body}\sim 5\times 10^{28}$body$^{-1}$s$^{-1}$,
$N_{CO}$ the number of bodies now evaporating CO around \bp\ must be
$N_{CO}=(M_{CO}\tau_{CO})/(Z_{\rm body}\mu_{CO})
\approx 6 \cdot 10^7 {\rm \ bodies}$.
This number is extremely large but unavoidable 
because CO is {\em observed}. 
These $\sim 10^7$--$10^8$~objects must be compared to the $10^8$--$10^9$
objects believed to be present between 30 and 100~AU from the Sun as 
the source of the Jupiter Family Comets. 
Anyway, the mass of parent bodies required by the evaporation process
(about one Earth mass, provided that 
some process is able to start its evaporation) is well below the mass 
needed to supply the \bp\ disk only 
by collision (30 Earth, Backman et al. 1995).

\subsection{$\beta\:$Pictoris a Transient Phenomenon ?}

It could be difficult to imagine that $\sim 10^8$~bodies have
always been active for $\sim 10^8$~years. 
$M_{CO}\times 10^8 {\rm years}\times \tau_{CO} = 20 M_{\rm Earth}$
of CO should have been evaporated !
It seems unlikely that this large number of bodies have 
been active since the birth of the system.
This gives evidence that either that \bp\ is very young 
or that it is a transient 
phenomenon. There is in fact no reason to believe that the \bp\ 
system was always as dusty as observed today. 
Of course, the idea that this disk is not transient is a consequence of any
model of collisional erosion from asteroid to dust. 
But with other scenarios, we can easily imagine that
a particular phenomenon occurred recently, and that the density 
of the \bp\ disk must be significantly smaller during the quiescent phase
of simple collisional erosion during which the density can be similar to 
the characteristic density of the more common Vega-like stars.

\section{The Asymmetry Problem}
\label{The Asymmetry Problem}
\label{Origin}

In contrast to a production by a set of collisional 
bodies at very large distances where planets have no influence, 
if the dust is produced by a narrow ring of orbiting evaporating
bodies, these bodies must be close to the star where the
planetary perturbations can be important.
In this case,
the asymmetry can simply be due to an eccentric orbit of
the perturbing planet. For instance, one major planet on an eccentric orbit
can cause a modulation of the precessing rate 
of the periastron of the OEBs. It is thus well-known that the distribution of
the perihelion of the asteroids in the Solar System is not
axisymmetric, and is closely related to the Jupiter longitude
of the perihelion (Kiang 1966). 
The density of asteroids with the same longitude 
of perihelion as Jupiter is thus $\sim 2.5$~times 
larger than that with periastron in opposite direction. 
This is simply because 
when the periastron of an asteroid is located at 180 degrees 
from the periastron of Jupiter, the precessing rate is quicker  
and the density is smaller.

\begin{figure}[tb]
\hspace{2.5cm}
\psfig{file=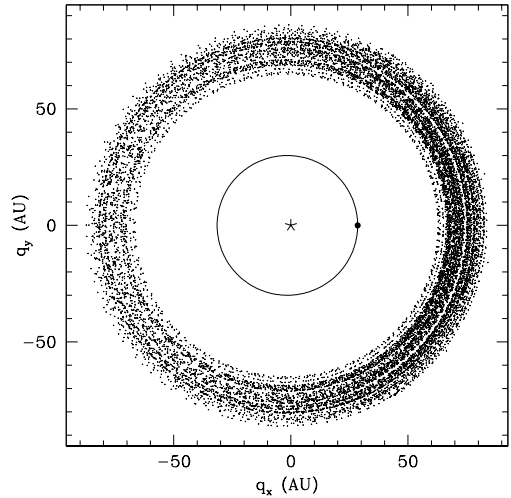,width=7cm}
\caption[]{Plot of the spatial distribution of the periastron of a set of 
bodies located between 70 and 90~AU and perturbed by a planet 
on eccentric orbit ($M_p= 3\cdot 10^{-4} M_{\odot}$, $e_p=0.05$, $a_p=30$~AU).
We see that the density of bodies with periastron in the direction of 
the planet periastron (black dot) is very large.
Moreover the periastron distances are also smaller. For these two reasons, 
the dust production by evaporation must be larger in this direction. 
If these bodies evaporate, they produce a dust disk which must be asymmetric.}
\label{dwdt2}
\end{figure}

Such an effect would obviously
cause an asymmetry in a disk \emph{if it is
produced by evaporation} of bodies with a distribution
of periastron perturbed in this way.
As the dust is mainly produced at the periastron of the parent bodies 
and principally observed during the apoastron, the part of the disk 
at 180 degrees from the
perturbing planet periastron could be more dense
(an example of such a situation is given in Fig~\ref{dwdt2}).

\section{Resonances with a Migrating Planet}
\label{migrating planet}

Possible origin of these OEBs, or more exactly the perturbations
necessary to explain their evaporation, have to be explored. 
Indeed, evaporation takes place only when 
a body is formed beyond a vaporization limit of a volatile and its 
periastron distance then decreases below this limit.

\subsection{The Formation of\ \ Uranus and Neptune}
\label{fernandez model}
 
To solve the problem of time scales for the formation
of the outer planet of the Solar System,
Fern\'andez (1982) suggested that  
the accumulation and scattering of a large number of planetesimals 
is the origin of the migration of the outer planets during their formation.
This migration is essentially due to the exchange of angular momentum 
between Jupiter and the proto-Uranus and proto-Neptune, via the accretion 
and gravitational scattering of planetesimals, the orbit of Jupiter loses 
angular momentum and shifts slightly inward, while those of Saturn, Uranus and
Neptune move outwards by several~AU. 
This model successfully explains the formation of the two outer 
planets of the Solar System, in short time scale 
($2\cdot 10^8$ to $3\cdot 10^8$~years),
their mass and their actual position (Fern\'andez \& Ip 1996).

The consequences of this scenario on the structure of the outer Solar System 
has been investigated by Malhotra (1993, 1995) who showed that this also
explain the particular orbit of Pluto with its large eccentricity and inclination,
and its resonance with Neptune. In short, Pluto was trapped in the 
orbital commensurability moving outward during the expansion phase of 
Neptune's orbit. 
The outward migration of Neptune can also 
explain the fact that numerous Kuiper belt objects are observed
in Pluto-like orbit in 2:3 resonance with Neptune (Malhotra 1995).
 
\subsection{Planet Migration and Perturbation on Parent Bodies.}
\label{migration}

With this in mind, it is interesting to evaluate the possible link
between the migration of outer planets and the \bp-like circumstellar disks
for which we know that the age is similar to the time scale of formation of
these planets ($10^{8}$~years is about the age of 
\bp\ and $\alpha$~PsA). 
Following Malhotra, we numerically investigate the consequence of the migration
of the planets in the Fern\'andez's scheme
on the dynamical evolution of the planetesimals, and their possible
trapping in resonant orbits which allow evaporation of frozen volatiles. 
For simplicity we consider only one outer massive planet 
supposed to suffer an exchange of orbital angular momentum 
as a back-reaction on the planet itself of the planetesimal scattering.
Of course, at least a second inner planet must be there.
Here, we consider only the principal
outer perturbing planet which is supposed to migrate 
because of a force equivalent to a drag force decreasing 
with time: $F_D \propto e^{-t/\tau}$, where $\tau$ is the characteristic
time of the migration.

In fact, if the migrating planet is moving inward, the planetesimals
are not trapped in the resonances. Their
semi-major axis remain unchanged and their eccentricities are 
only slightly increased. Consequently, the decrease of the periastron
distance is too small to allow the volatiles to evaporate.

On the contrary, if the outer planet is moving outward, as found 
in the models of Fern\'andez \& Ip (1996),
a fraction of bodies can be trapped in resonances.
Their semi-major axis and eccentricities increase significantly
and the 
net 
result is a decrease of their periastron.
This can start the evaporation of the trapped bodies.

\begin{figure}[tb]
%\hspace{2.5cm}
\psfig{file=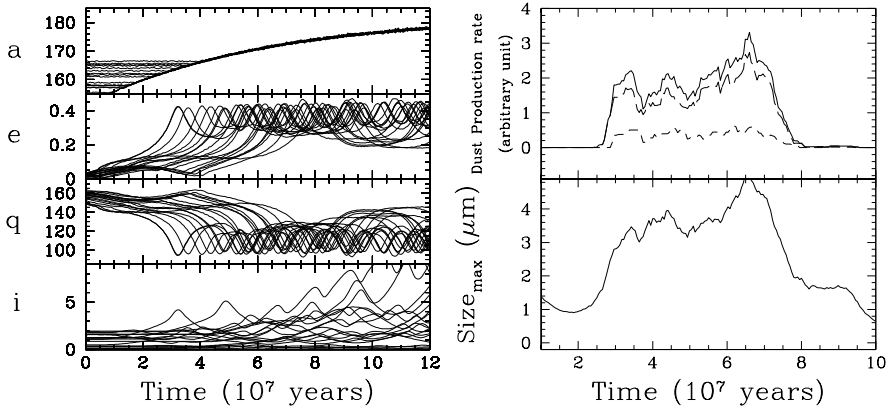,width=\columnwidth,angle=-90}
\caption[]{Left panel. Plot of the orbital parameters semi-major axis ($a$), 
eccentricity ($e$), periastron ($q$) and inclination ($i$) of 19 bodies trapped
in the 1:4 resonances with a migrating planet.

We see that although the semi-major axis of the bodies
increase, their periastron decrease. These bodies can
start to produce dust by evaporation.\\

Right panel.
Plot of the dust production as a function of time for evaporating 
bodies trapped in 1:4 resonance with a migrating planet.
This is the dust production for grains larger than 2$\mu$m, 
because smaller grains 
are supposed to be quickly expelled by the radiation pressure.
The production of dust starts when the periastron distance of the
parent bodies is small enough for the CO production rate to
allow ejection of grains larger than 2$\mu$m.
Then, it stops when the parent bodies are exhausted and 
have no more volatile. 
Consequently, the dust production is transient
and is large only when the bodies trapped in the 
resonances are entering in the evaporation limit. 

The production rate is also not axisymmetric. As in Sect.~\ref{Origin},
we see that the production is 
larger in the direction of the periastron of the perturbing planet 
(long dashed) than in the opposite direction (short dashed).\\

The right bottom panel gives the corresponding maximal size of grains ejected 
from the bodies by the evaporating gas.
Because the periastrons are still
larger than 100~AU, the evaporation produces only small particles.
In this simulation, the maximal particle size is around 4$\mu$m
as expected to explain the slope $\alpha\sim 4$ observed in the \bp\ case. }
\label{pna2t}
\end{figure}

We have tested several configurations of outward migration
and have evaluated the effect on planetesimals in the zones swept
by first order resonances. 
The 1:2 and 1:3 resonances does not allow to explain the observed
characteristics of the \bp\ disk.
The 1:4 resonance give the most interesting results
(Fig.~\ref{pna2t}). The trapping has been found to 
be efficient if the mass of the planet
is $M_p \ga 0.5 M_J$, where $M_J$ is the mass of Jupiter, 
and if the migration rate is 
$\tau \ga 5\cdot 10^{7}$~years. 
With these conditions, the periastron of trapped bodies significantly 
decreased. A significant increase 
of the inclination has also been observed after few $\tau$ as well as 
a large asymmetry in the distribution of the periastrons longitude.
The 1:5 resonance is efficient in trapping
only if the parameters of the migrating planet are extreme with 
$M_p\ga M_J$, $e_p \ga 0.1$ and $\tau \ga 5\cdot 10^{7}$~years.

\subsection{Asymmetry and Opening Angle}

If the bodies are trapped in a resonance with a planet on eccentric
orbit, there can be an asymmetry in the distribution of the 
periastron as already seen in Sect.~\ref{Origin}. 
For example, the Fig.~\ref{pna2t} gives the dust production rate by
the bodies trapped in the 1:4 resonance with a planet on an eccentric
orbit ($e_p=0.05$). The production rate is larger 
in the direction of the periastron of the planet
than in the opposite direction.
The disk thus produced must be asymmetric with a larger density
in the direction of the apoastron of the migrating planet.

From Fig.~\ref{pna2t}, we
also conclude that the production
of dust can take place with 
the inclination of the parent bodies 
larger than the initial inclination, up to several degrees.
Moreover, with several giant planets, the precession
of the ascending nodes can produce an additional 
increase of the parent bodies inclination. 
In all cases, this migrating and resonance 
trapping process gives a large increase in the inclinations
and consequently a large opening angle of the associated dust disk.

\section{Conclusion}
\label{conclusion}

Collisions and evaporation are the two main processes believed to be able to
supply disks like the \bp\ one. 
These two processes 
are not exclusive.
However, the \bp\ disk is more likely produced by the evaporation process. 
The CO and \ci\ gas detected with HST definitely shows that evaporation 
takes place around \bp , even if its consequence on dust replenishment 
in comparison to the collisional production is still a matter of debate.
The dust spatial distribution with the slope of the power law, 
and the central hole can be explained 
by the characteristic distances of evaporation. Finally, 
the asymmetry at large distances can easily be explained 
in evaporating scenarios because
the parent bodies are maintained close to the star 
where planets' influences are important. The asymmetry 
is then simply a consequence of the non-axisymmetry of the perturbation 
by planet(s) on eccentric orbits.

We have shown the possibility that bodies trapped in resonances with a 
migrating planet can evaporate. The large number of CO
evaporating bodies is explained by
a transient evaporation during a short period.

From another point of view, if the migration of the outer planets 
took place in the Solar System, 
why not around other stars ?
This is in fact a simple consequence of the presence of a forming
planet inside a disk of residual planetesimals.
Here we have explored a new consequence of this migration of a forming planet.
Some planetesimals can be trapped in resonances, enter inside
evaporation zone and finally become parent bodies 
of \bp -like disks. 
In short, as a direct consequence of the formation
of outer planets in the Fern\'andez's scheme, evaporation of Kuiper belt-like
objects around bright stars can be expected to be common. 
This allows us to look at the circumstellar disks around main sequence 
stars as a possible signature of outer planet formation.

\begin{acknowledgements} 
I am particularly grateful to J.M. Mariotti and D. Alloin for organizing this
very interesting and fruitful meeting.
\end{acknowledgements}

\end{document}